\documentclass[final]{svjour3}
\usepackage{graphicx}
\usepackage{rotating}
\usepackage{amssymb}
\usepackage{mathptmx}
\usepackage[numbers]{natbib}

\makeatletter
\journalname{Journal of Low Temperature Physics}

\usepackage[export]{adjustbox}
\usepackage{siunitx}
\usepackage{mhchem}
\usepackage{multicol}
\usepackage{upgreek}

\bibpunct{}{}{,}{s}{}{,}

\newcommand{\iprod}[1]{\left\langle #1\right\rangle}
\newcommand{\pth}[1]{\left( #1\right)}
\def\degr{^\circ}
\def\nqp{n_{\mathrm{qp}}}

\begin{document}

\newcommand{\hdblarrow}{H\makebox[0.9ex][l]{$\downdownarrows$}-}
\title{Atomic Layer Deposition Niobium Nitride Films 
for \\High-Q Resonators}

\author{Calder Sheagren$^1$\and Peter Barry$^2$\and Erik Shirokoff$^1$\\\and Qing Yang Tang$^1$}

\institute{$^1$ Kavli Institute for Cosmological Physics, University of Chicago,  Chicago, IL, 60637, U.S.A.\\
$^2$ Argonne National Laboratory, Lemont, IL, 60439, U.S.A.\\
\email{calderds@uchicago.edu}}

\maketitle

\begin{abstract}

Niobium nitride (NbN) is a useful material for fabricating detectors because of its high critical temperature and relatively high kinetic inductance. In particular, NbN can be used to fabricate nanowire detectors and mm-wave transmission lines. When deposited, NbN is usually sputtered, leaving room for concern about uniformity at small thicknesses. We present atomic layer deposition niobium nitride (ALD NbN) as an alternative technique that allows for precision control of deposition parameters such as film thickness, stage temperature, and nitrogen composition. Atomic-scale control over film thickness admits a high degree of uniformity for films 4-30 nm thick; control over deposition temperature gives rise to growth rate changes, which can be used to optimize film thickness and critical temperature. In order to characterize ALD NbN in the radio-frequency regime, we construct single-layer microwave resonators and test their performance as a function of stage temperature and input power. ALD processes can admit high resonator quality factors, with $\geq 43\%$ of resonators above $Q_i = 10^6$, which in turn increase detector multiplexing capabilities. Furthermore, we find critical temperatures in the range of $7.5$K to $10.9$K that vary as a function of cycle count and deposition temperature. 


\end{abstract}

\section{Introduction}
Thin niobium nitride (NbN) films have a range of potential applications in the development of superconducting detectors for mm-wavelength astronomy, photon and particle detection, and terrestrial applications.  The material's superconducting transition temperature ($T_c \geq 11\si{K}$) and microwave properties make it a promising material for low loss transmission lines and planar antenna components at submm-wavelengths shorter than is possible with Nb films. We are developing thin ($\sim 30\,$nm) ALD films with the goal of producing extremely uniform submm-wavelength band-defining on-chip features.

The high kinetic inductance, high normal resistivity, and internal quality factors ($Q_i$) exceeding $10^5$ of NbN films make it an appealing candidate for microwave resonators and devices which depend on non-linear kinetic inductance. ALD-deposition enables uniform and ultrathin ($<\SI{10}{nm}$) films which are well suited for these devices, which include Kinetic Inductance Parametric Up-Converters (KPUPS) \cite{kpups}, parametric amplifiers \cite{paraamps, paraamps2}, variable delay lines \cite{lk-delay}, and nanowire photon detectors \cite{nanowire, nanowire2}.  

Succinctly, we want to investigate the properties of ALD NbN in the ultrathin film regime ($t<10$nm) and thin film regime ($t\sim 30$nm) for use in nanofabricated superconducting detectors. 

\section{Resonator Fabrication}

\begin{figure}
\includegraphics[width=.3\textwidth,angle=270,valign=m]{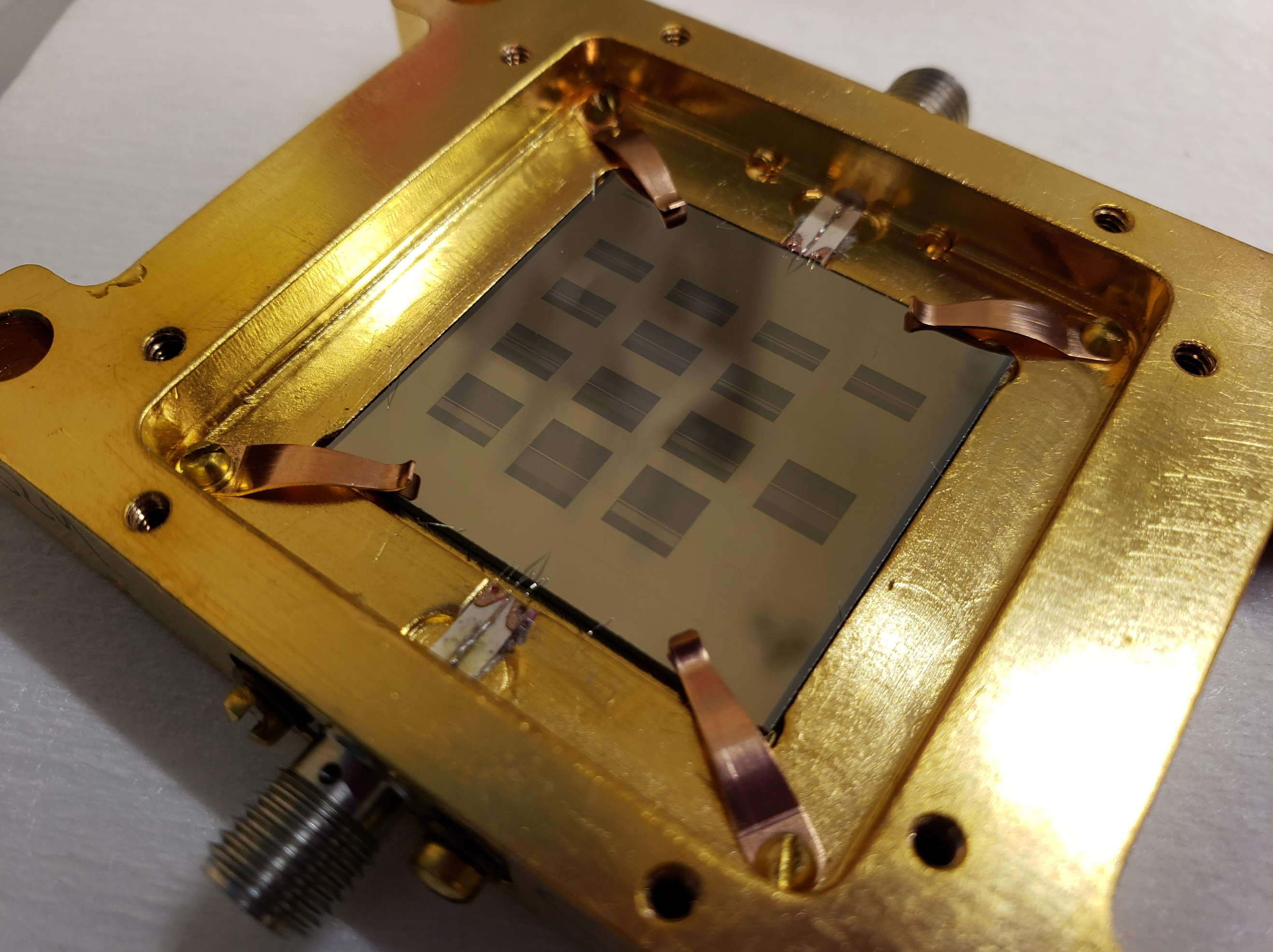}\hspace{.75cm}
\includegraphics[width=.56\textwidth,valign=m]{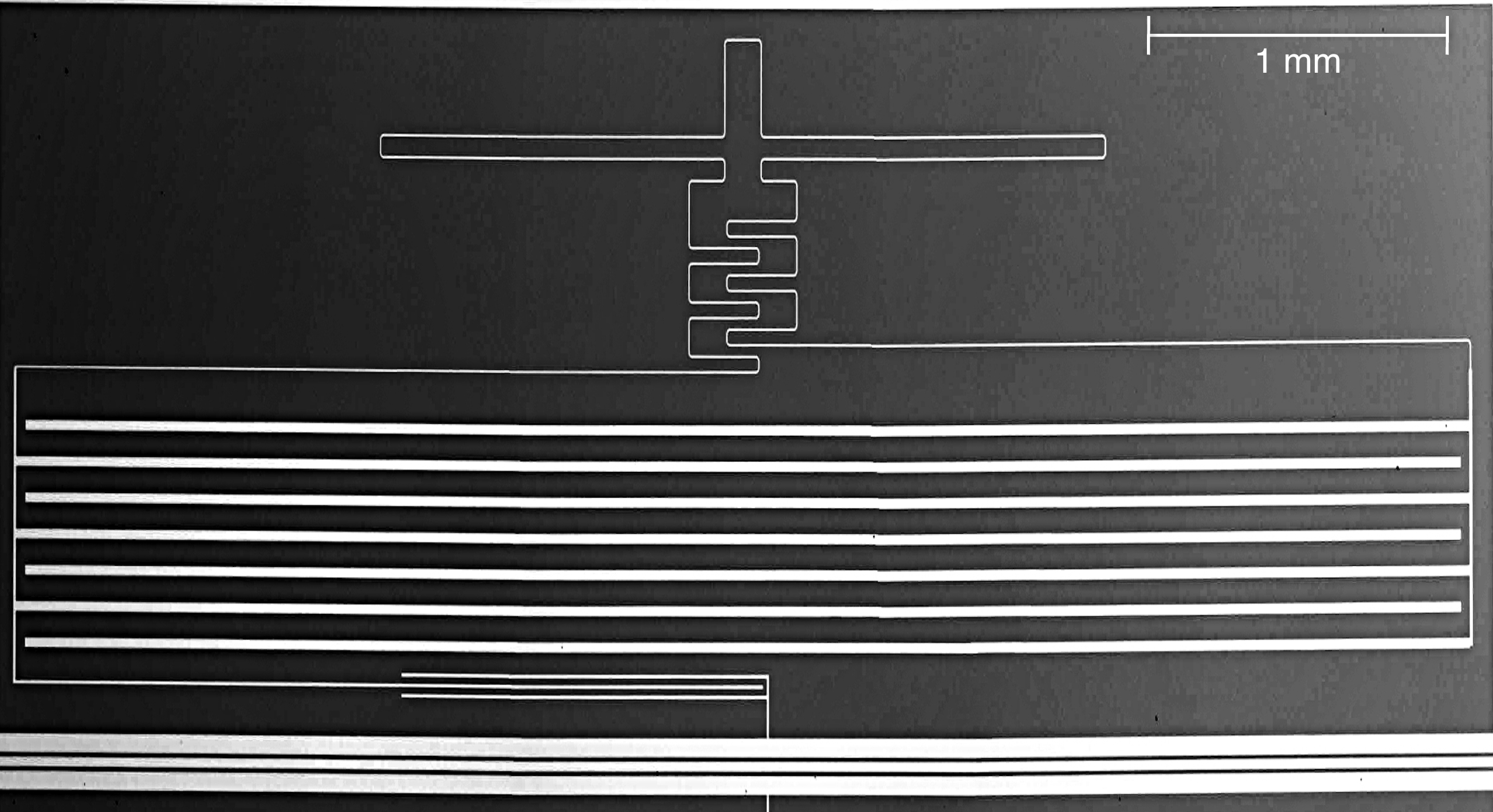}

\caption{(Color figure online.) (Left) wire-bonded NbN resonators in sample box. (Right) Composite microscope image of single resonator.}
\label{machinery}
\end{figure}
To construct microwave resonators from ALD NbN, we follow a standard single-layer fabrication process adapted for NbN processing, making use of the equipment at the Pritzker Nanofabrication Facility at the University of Chicago.  In particular, we fabricate planar lumped-element resonators with interdigitated capacitors and inductors similar to mm-wave optical coupled devices; feature sizes are on the order of microns, and the feedline width is $s_{\text{cpw}} = 16\upmu $m. Our devices have approximately 30 resonators per chip, and device pictures and schematics can be found in Fig. \ref{machinery}. 

\subsection{Deposition}

Before deposition, we perform a 10-minute solvent clean on our substrate, High-$R\, \iprod{100}$ Si, before immediately transferring it into the Ultratech Fiji Plasma-Assisted ALD machine. Before the deposition begins, the sample is held in a vacuum at high temperatures $(250\degr\si{C}\leq T\leq 300\degr\si{C})$ for between 6-60 minutes while the machine stabilizes, effectively acting as a vacuum bake. Furthermore, once the sample is in the ALD machine, we do not break vacuum until the deposition is complete, giving an optimally dehydrated environment.  \\

One Plasma-Enhanced ALD (PEALD) cycle consists of treating the substrate with the precursor, (tert- butylimido)-tris (diethylamino)-niobium (TBTDEN), removing the precursor from the chamber, lighting the plasma with $P=300\si{W}$, waiting, turning off the Argon plasma, and removing the excess Ar from the chamber. Furthermore, since Nb has a low vapor pressure, we use three Ar boost cycles to improve chamber precursor concentration. One ALD cycle creates approximately one atomic monolayer of metal, and we repeat this process for a given cycle count to achieve the desired thickness. Also, the stage temperature is highly connected to the growth rate of PEALD films\cite{ald-nbn}, so varying the stage temperature and thickness give a large parameter space to explore. 

\subsection{Lithography and Etch}
After the deposition, we perform an 8-minute vacuum dehydration bake before spinning AZ1512 photoresist at 3500 RPM. We then soft-bake the wafer at $90\degr\si{C}$ for one minute and expose the desired pattern on the Heidelberg MLA150 Laser Writer at a dose of $125\si{ mJ}$ on the 375nm laser. We do not rely on complex E-beam lithography since our minimum feature size is $>2\si{\upmu m}$. Afterwards, we hard-bake the wafer for 1 minute at $110\degr \si{C}$ and develop the exposure in AZ1:1 developer for 1'30 before checking the exposure quality under a microscope. 

Once we are satisfied with the exposure quality, we perform an \ce{O2} descum to remove a small amount of photoresist and use a Fluorine-based Inductively-coupled Plasma etch process to remove the NbN. This etch uses 40sccm \ce{CF4}, 10sccm \ce{CHF3}, and 10sccm Ar, and we etch the sample between 5-10 minutes depending on the thickness. Once we are satisfied we have etched through the NbN, we use an Oxygen ash to remove the hard photoresist crust before soaking the sample in heated NMP for at least 2 hours to remove the remainder of the photoresist. We finally dry and dice the chips before we mount them to test.
\section{Measurements}
\subsection{Material Properties}

\begin{figure}
\centering
\includegraphics[width=.5\textwidth , valign=m]{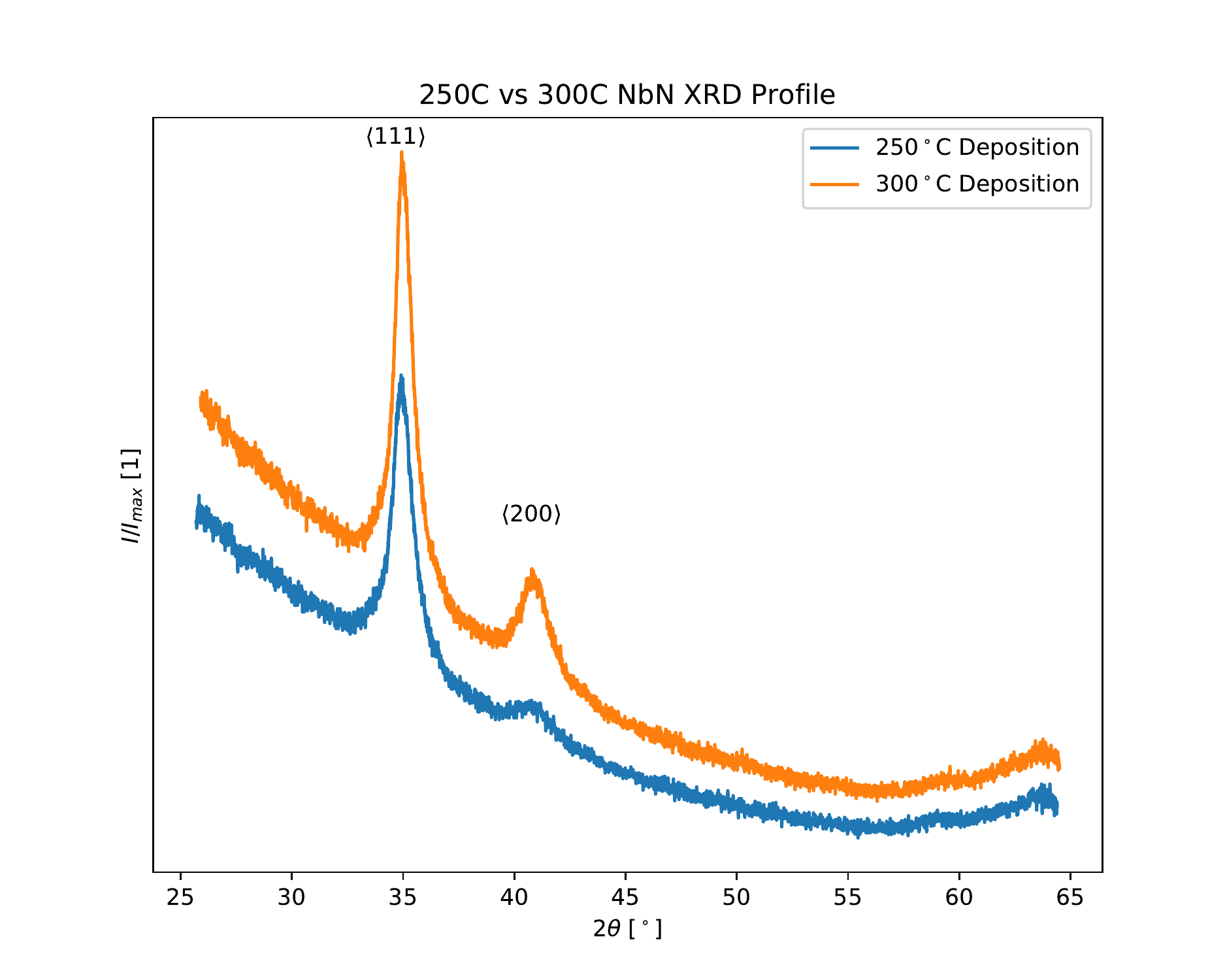}
\includegraphics[width=.37\textwidth , angle=270,valign=m]{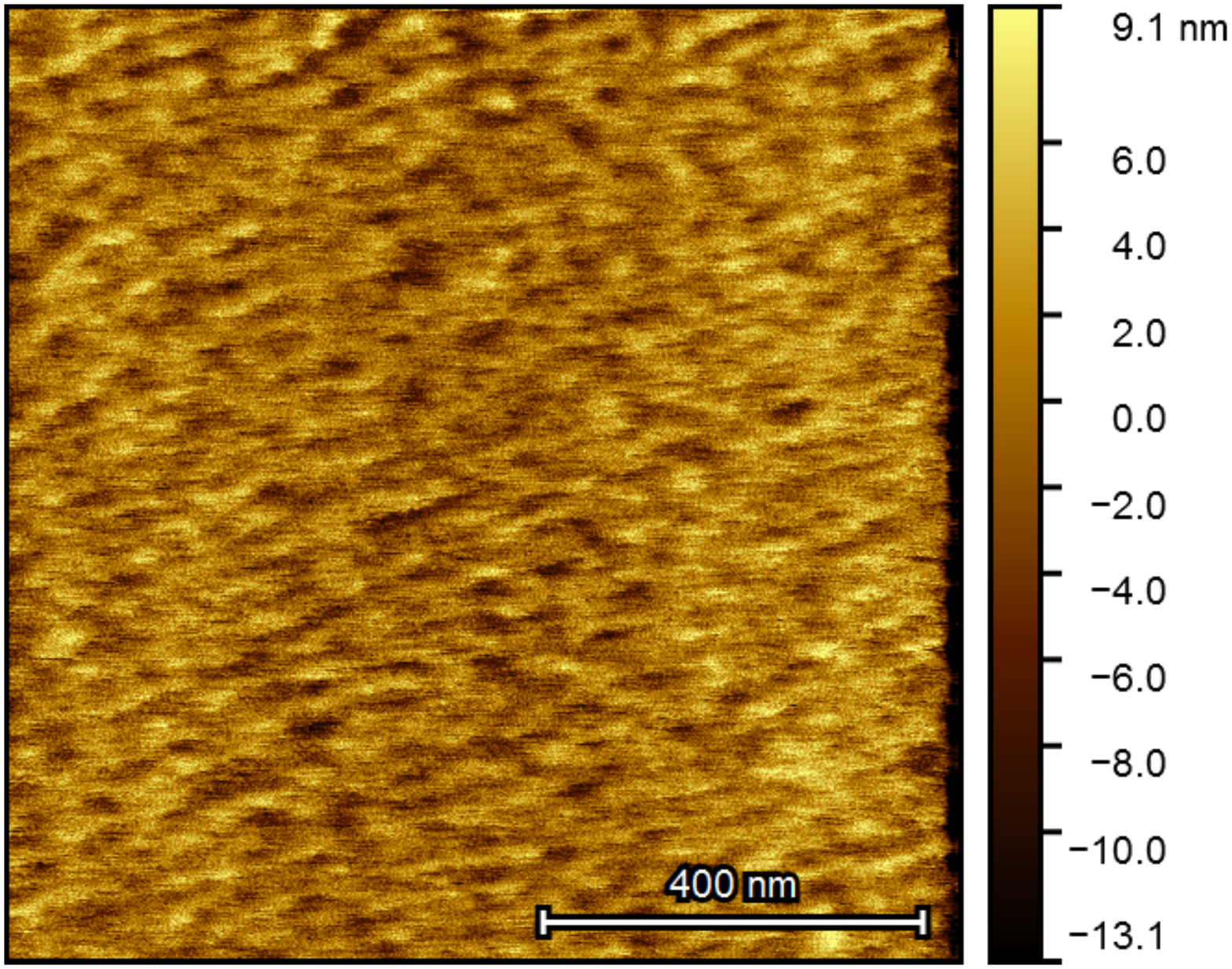}

\caption{(Color figure online.) (Left) XRD profile for thin NbN films deposited at 250$\degr$C and 300$\degr $C. (Right) Atomic Force Microscopy image of thin NbN sample.}

\label{fig 1}
\end{figure}

We measured film thickness using ellipsometry and obtained a growth rate of $.51 \pm .05\,\AA/$cycle for films deposited at $250\degr $C and $.62\pm .05\,\AA/$cycle for films deposited at $300\degr $C. 
These values are consistent with Sowa et. al.\cite{ald-nbn}, but are larger than those reported by Linzen et. al\cite{linzen}, who used a higher deposition temperature of 350$\degr$C.  

X-ray diffraction (XRD) patterns were obtained using a Bruker D8 diffractometer with a Cu $K_\alpha$ X-ray source operating at 40 kV and 40 mA and a Vantec 2000 area detector. The XRD profiles for films deposited at $250\degr $C and $300\degr$ C are shown in Fig. \ref{fig 1}. To find the ratio of the various crystal structures, we subtracted out the baseline and compared the areas under the peaks corresponding to each crystal orientation. We find a ratio of $\iprod{111}:\iprod{100}=3.1\pm .5$ for films deposited at $250\degr$C and a ratio of $\iprod{111}:\iprod{100}=2.8\pm .5$ for films deposited at $300\degr$C. 

We also used Atomic Force Microscopy on a Bruker Atomic Force microscope to determine the surface roughness, which we found to be $3.0\pm .1$ nm for thin films deposited at $250\degr \si{C}$ and $3.6\pm .3$ nm for thin films deposited at $300\degr\si{ C}$. The open-source program Gwyddion was used for image manipulation, including level correction and horizontal `scar' correction among other techniques. Granular structures were not observed when measuring films thicker than 300 cycles at a $1\si{\upmu m}^2$ field of view. 

\subsection{DC Measurements}

\begin{figure}

\includegraphics[width=.5\textwidth]{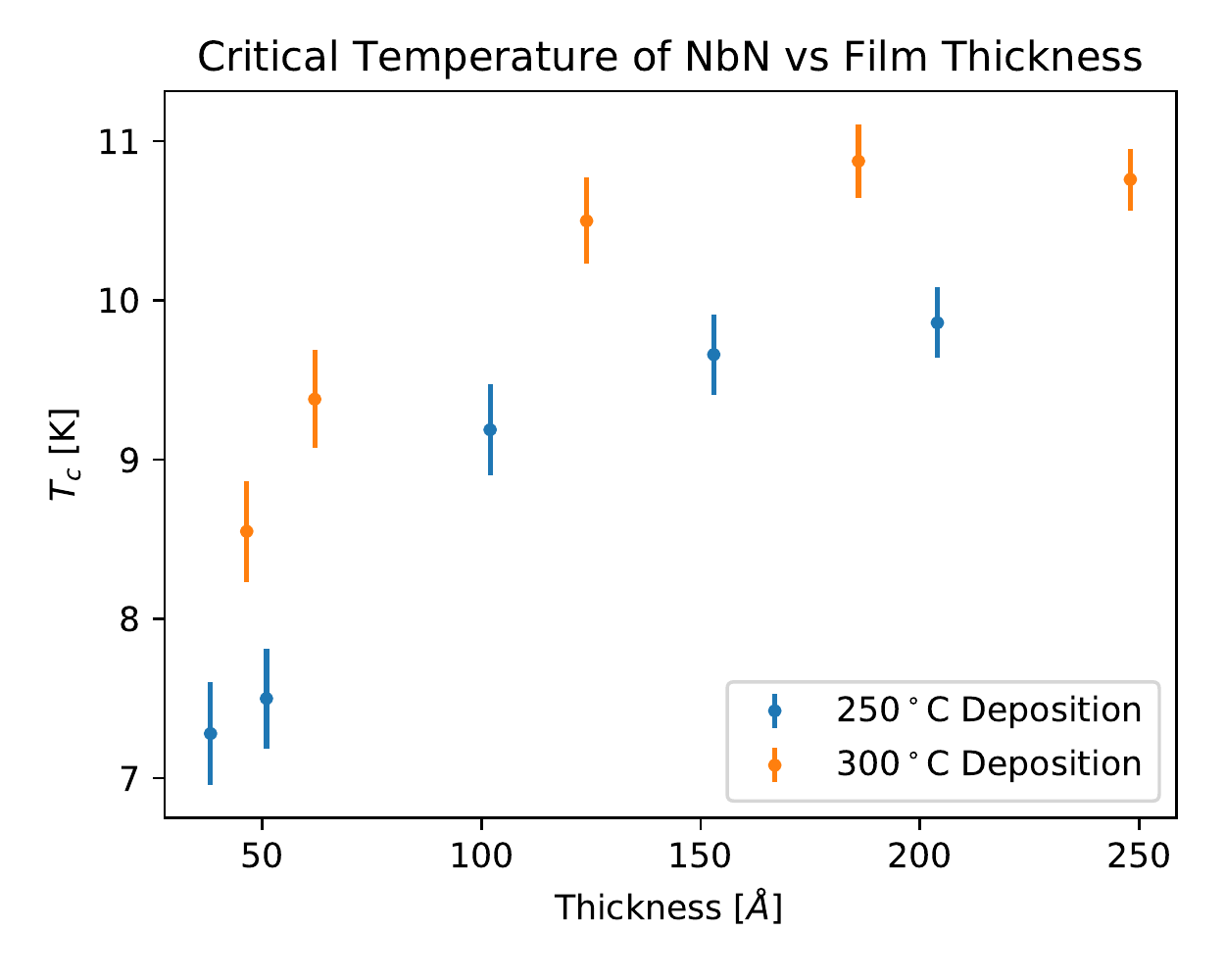}
\includegraphics[width=.5\textwidth]{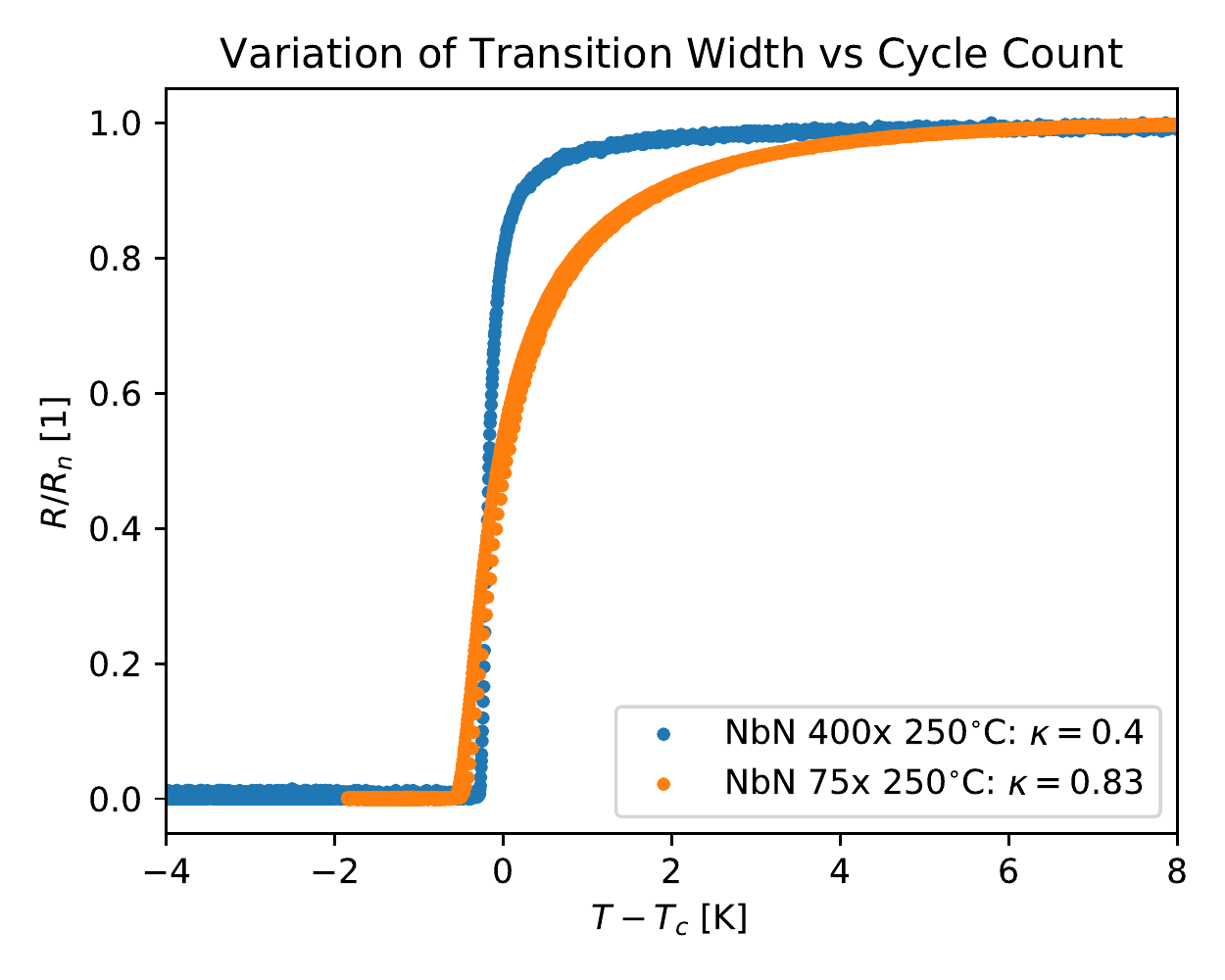}

\caption{(Color figure online.) (Left) $T_c$ of NbN samples as a function of thickness, with deposition temperature noted.   (Right) Resistance vs Temperature curves for two NbN films (Blue: 400 cycles, $t \sim 20$nm;  Orange: 75 cycles, $t\sim 4.7$nm) deposited at the same temperature with different cycle counts.}
\label{tw-tc}
\end{figure}

Using an AC resistance bridge, we measure the sample resistance of either unpatterned wafer samples or patterned rectangular van Der Pauw features to obtain the film resistivity and DC critical temperature. We have not seen films less than 75 cycles superconduct, and the highest critical temperature we observed was $T_c = 10.87$K, much lower than has been observed with other ALD NbN processes\cite{ald-nbn,linzen}. Additionally, we notice a 12\% variation in critical temperature when measuring different samples from the same wafer or deposition with a spatial separation of 2-4cm. 

Most of our samples were unpatterned bare films, -- we wanted to quickly test for superconductivity and $T_c$ -- so we present resistivity measurements for the van der Pauw measurements only. We found that for $N_{\text{cycles}}= 300$, films deposited at 250$\degr$C had a resistivity of 160$\upOmega/\square$, and films deposited at 300$\degr$C had a resistivity of 78$\upOmega/\square$ . Additionally, we found $\rho_n = 301\upOmega/\square$ for a 75 cycle film deposited at 300$\degr$C. 

Resistance vs temperature data was collected in the temperature range of 1K to 15K, and data was fitted to an ad hoc fitting function 
\begin{equation}
R(T)={2R_n\over \pi}\tanh\pth{\kappa(T-T_c)} + {R_n\over 2}, 
\end{equation}
designed to select $T_c$ as the point where the resistance is ${R_n/2}$, where $R_n$ is the normal resistance. Additionally, $\kappa$ is a parameter that lends insight into the transition width; we observe that the transition width increases as film thickness decreases, in contrast to Linzen et. al.\cite{linzen}. In particular, increased transition width for thinner films is thought to imply an increase in granularity as film thickness decreases. Fig. \ref{tw-tc} shows two normalized $R$ vs $T$ plots for different film thicknesses. 

\begin{figure}
\begin{center}
\includegraphics[width=\textwidth]{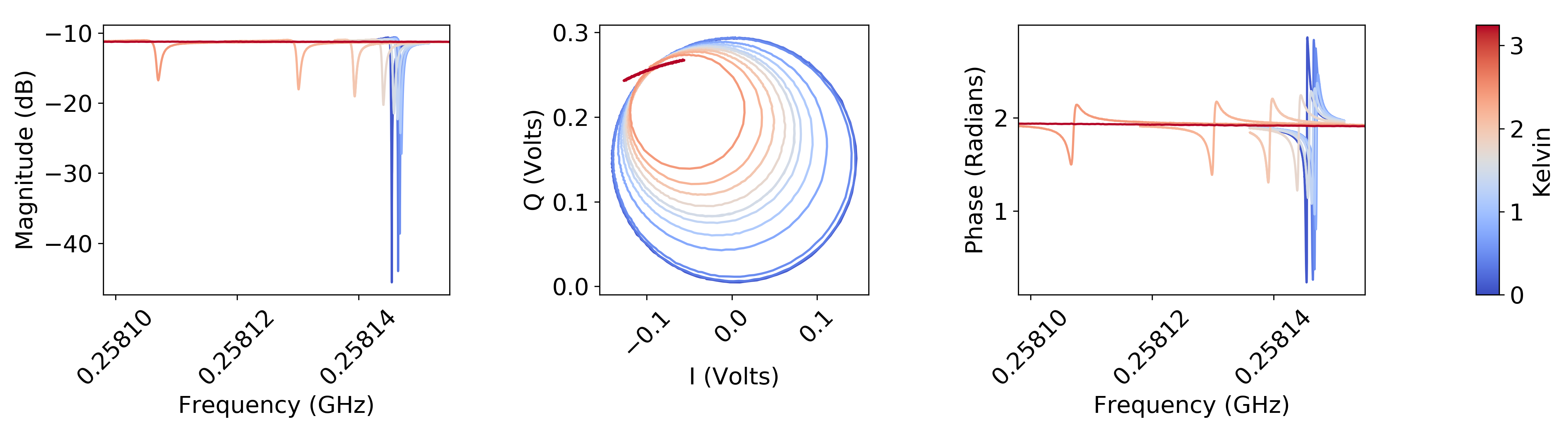}
\end{center}
\caption{(Color figure online.) Resonator behavior for a 300 cycle film deposited at 250$\degr$C, $t\sim 15.3$nm, as a function of cryostat stage temperature. (Left) $|S_{21}|$ vs frequency. (Center) $\mathrm{Im}(S_{21})$ vs $\mathrm{Re}(S_{21})$. (Right) Phase vs frequency.}
\label{s21-t}
\end{figure}

\subsection{Resonator Fits}\label{analysis}

To test the resonators at cryogenic temperatures, we use a BlueFors Helium dilution cryostat with appropriate coaxial wiring, filtration, and attenuation both at cryogenic and room temperatures. We measure the profile of each resonator at a range of readout powers and temperatures with a vector network analyzer.

Raw resonator data was fitted to the equation 
\begin{equation}
S_{21}(f)=1-{Q_r\over Q_c}\cdot {1\over 1+2jQ_r\cdot\pth {f-f_0\over f_0}}+\text{baseline},
\end{equation}
where $f_0$ is the resonant frequency, $Q_i$ is the internal quality factor, $Q_c$ is the coupling quality factor, and $Q_r^{-1}=Q_i^{-1}+Q_c^{-1}$. We used the \verb|scraps| python package\cite{Carter2016} to initially fit and process the data; Fig. \ref{s21-t} shows sample resonator data as a function of temperature. Afterwards, we perform additional Mattis-Bardeen fitting\cite{Gao2008} to extract RF parameters. In particular, we use the relations 
\begin{equation}
 {\delta f_0\over f_0}={\alpha_k\gamma\over 2}{d \ln(\sigma_2)\over d\nqp }\delta \nqp\qquad\delta \pth{1\over Q_i}={\alpha_k\gamma\over \sigma_2} {d \sigma_1\over d\nqp}\delta \nqp
\end{equation}
to fit for the kinetic inductance fraction $\alpha_k$ and gap energy $\upDelta_0$,  and use the relation $\upDelta_0 = 1.76kT_c$ to obtain $T_c^{(RF)}$. Note that we are in the local limit, so $\gamma={1\over 2}$, and $\sigma=\sigma_1+j\sigma_2$ represents the complex conductivity. This paper uses the convention that `$d$' represents analytic derivatives, and `$\delta$' represents computer-calculated differences. These relations allow us to fit  $df/ f_0$ vs $T$ and $1/Q_i$ vs $T$ as calculated from data to obtain values for $\alpha_k$ and $T_c^{(RF)}$.

To find the kinetic inductance ($L_k$) values for the NbN films, we compare the NbN devices with previously tested Al devices with the same circuit design (feedline width $s_{\text{cpw}}=16\,\si{\upmu m},\,$ thickness $ t=50\si{nm}$), using the relation 
\begin{equation}\label{f0-Lk}
f_0={1\over 2\pi\sqrt{(L_k+L_g)C}},
\end{equation}
where $L_g$ is the geometric inductance, and $C$ is the capacitance. We can estimate\cite{gao-thesis} $\alpha_k\approx .15$ for the Al chips and use that value to solve for $L_g,C$ for the circuit. Plugging these values into Eq. \ref{f0-Lk}, we extract $L_k=5.4\pm 2\,\si{pH}/\square$ for a 300 cycle film deposited at $250\degr $C, and $L_k=1.7\pm .5\,\si{pH}/\square$ for a 300 cycle film deposited at $300\degr $C.

Quality factors for NbN chips were found to be above $10^5$ for nearly all resonators, with $75\%$ above $10^6$ for a 250C film and $43\%$ above $10^6$ for a 300C film. Additionally, we measured different devices from the same 4" wafer and saw average quality factors differ by a factor of 10; we also observed a predictable rolloff in quality factors as the stage temperature increases. Quality factors for resonators as a function of deposition temperature are shown in Fig. \ref{q,rftc}. 

\begin{figure}

\includegraphics[width=.52\textwidth,valign=m]{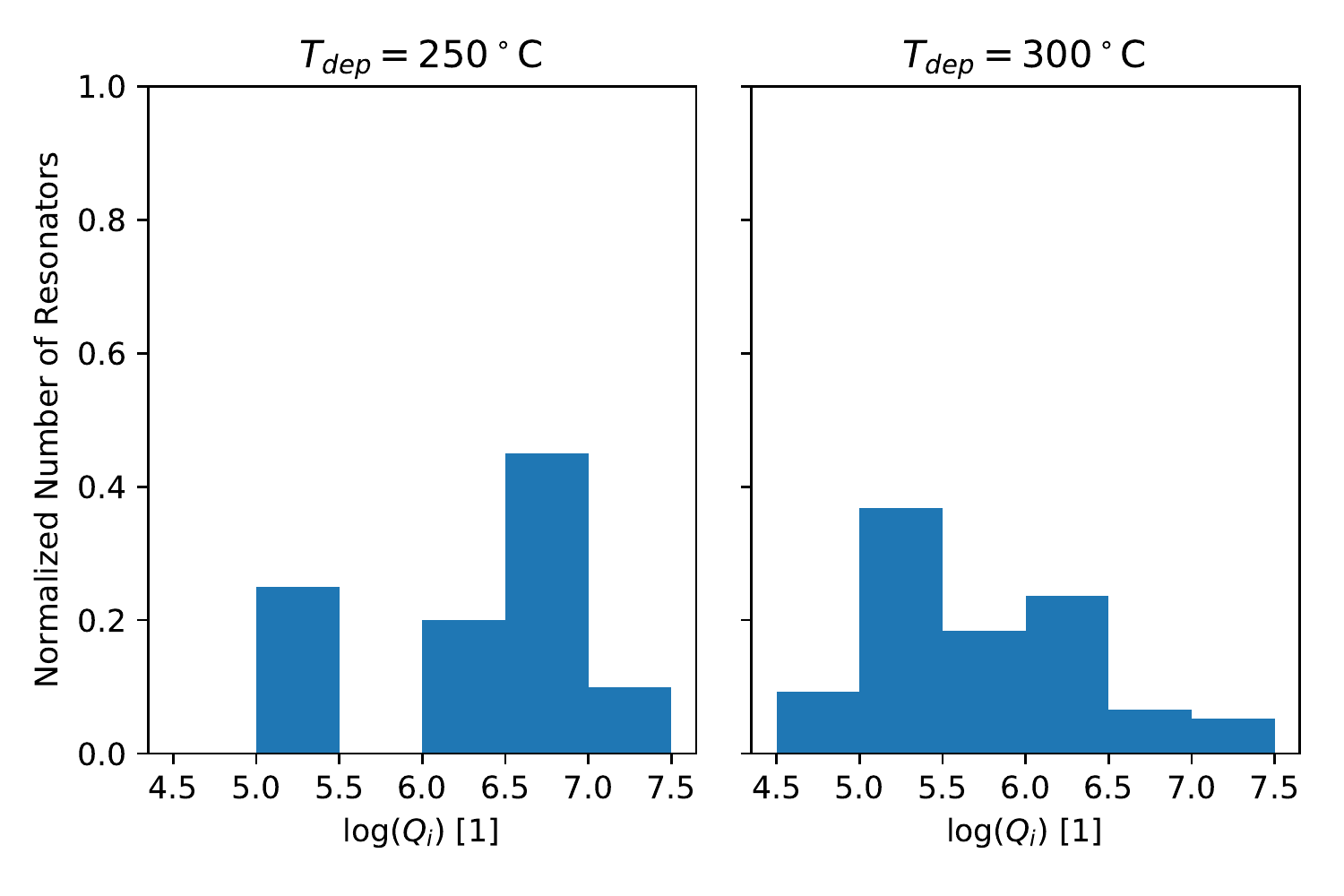}
\includegraphics[width=.45\textwidth,valign=m]{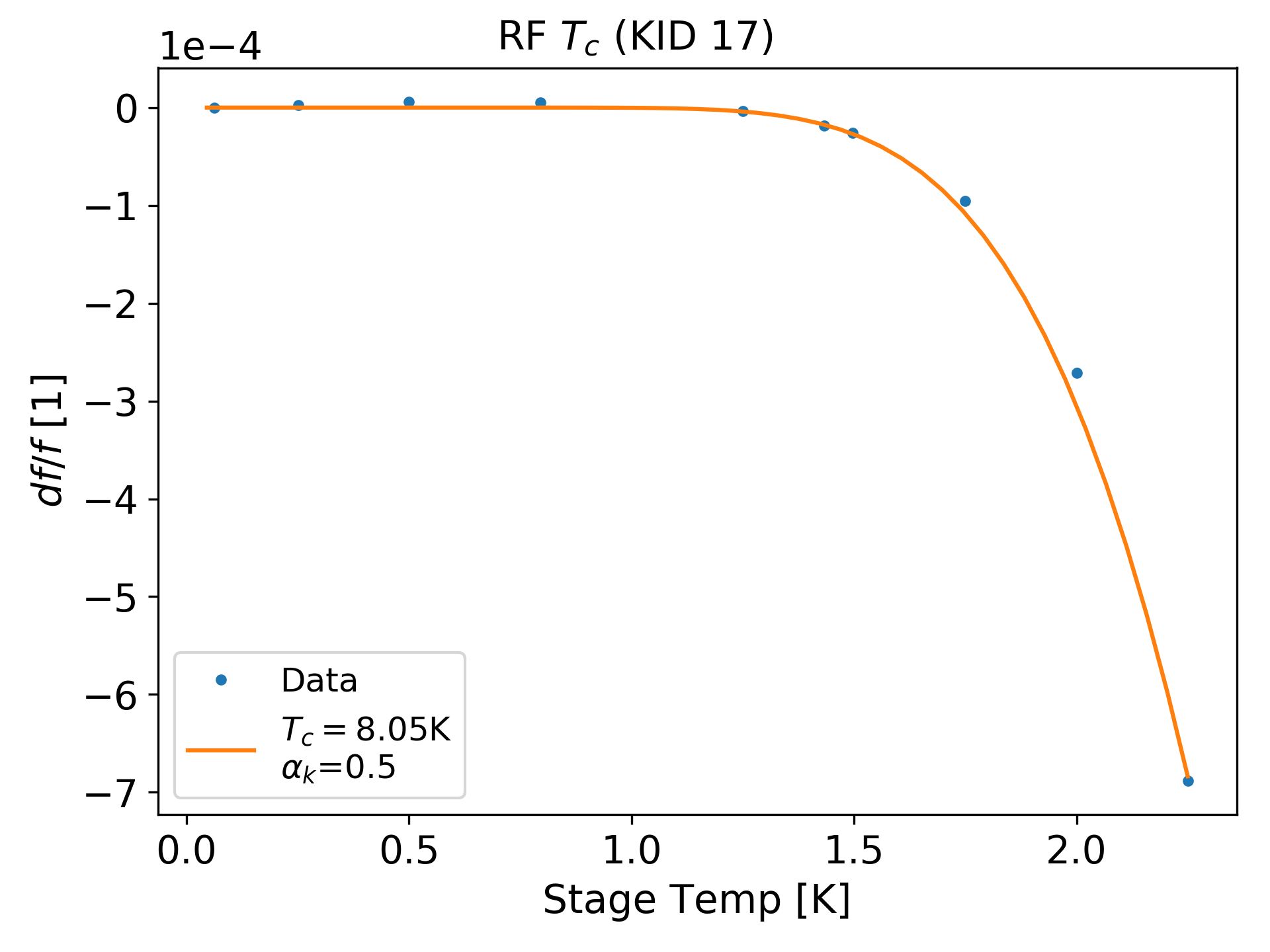}

\caption{(Color figure online.) (Left) $Q$ factors for resonators as a function of deposition temperature, with $T_{\text{stage}}=300$mK. These $Q_i$ values were obtained with a formal fit for resonators with $Q_c$ values between $10^4$ and $10^6$. (Right) Resonator $df/f$ vs $T$ behavior for a 75 cycle film deposited at 300$\degr$C, $t\sim 4.6$nm.}
\label{q,rftc}
\end{figure}

We fitted for $T_c^{(RF)}$ using the fractional frequency shift as a function of stage temperature, and found $T_c^{(RF)}\sim 11$K for 250$\degr$C films, and $T_c^{(RF)}\sim 12$K for 300$\degr$C films. 
\section{Discussion}
As discussed in Section \ref{analysis}, $Q$ values for ALD NbN resonators are high and comparable to those offered by sputtered NbN films. While we have not achieved the high  $T_c\sim 13-15$K\cite{faustin-nbn, spiral-mkid} of some sputtered films, quality factors are higher than as reported with a spiral-design NbN MKID\cite{spiral-mkid} and are comparable to NbN CPW resonators\cite{faustin-nbn}. Furthermore, ALD film thicknesses are much smaller than those reported for high-$T_c$ sputtered films; we hope to investigate more resonators tested from thinner films in the future.

\section{Conclusion}

In conclusion, test ALD NbN resonators were fabricated, showing high $Q$ values and a moderate $L_k$. DC samples showed a characteristic increase in $T_c$ with thickness, albeit reaching a maximum of around 10.87K, which is lower than the bulk value.  

In the future, we hope to increase the DC $T_c$ further by modifying deposition parameters, surface preparation, and substrate type. In particular, modifying the Ar pressures during the deposition and changing the substrate lattice to sapphire or $\iprod{111}$ Si may have a large effect on the film. Based on these promising initial results, we will pursue additional parameter variations and multilayer test devices produced from these films.\newpage

\begin{acknowledgements}
This work was supported by the National Science Foundation under grant no. AST-1554565, the Kavli Institute for Cosmological Physics, and the Pritzker Nanofabrication Facility of the Institute for Molecular Engineering at the University of Chicago, which receives support from SHyNE, a node of the NSF's National Nanotechnology Coordinated Infrastructure (NSF NNCI-1542205). 
The authors would like to thank Alexander Anferov and David Schuster for their generous help in this project and Kirit Karkare for his help in revising this manuscript. 
\end{acknowledgements}

\bibliographystyle{jltp}
\bibliography{nbn}

\end{document}